\documentclass [11pt]{article}
\usepackage{setspace}
\usepackage{endnotes}
\usepackage{fullpage}
\title{The behaviour of rods and clocks in general relativity, and the meaning of the metric field}
\author{ Harvey R. Brown\\
Faculty of Philosophy, University of Oxford\\ 10 Merton Street, Oxford OX1
4JJ, U.K.\\{\em harvey.brown@philosophy.ox.ac.uk}}
\date{}

\begin{document}
\maketitle

\textit{In this
view what general relativity really succeeded in doing was to eliminate geometry from physics.} James L. Anderson (1999)

\section{Introduction}

Why is the $g_{\mu\nu}$ field commonly assigned \textit{geometrical} 
significance in general relativity theory? Why is it often regarded the fabric of space-time itself, or the `arena' of dynamical processes? The standard explanation suggests that in part,
this is because rods and clocks survey it (if only approximately) in a way that does not depend on their constitution. Here is how Clifford Will put the point:
\begin{quotation}
The property that all non-gravitational fields should couple in the same manner to a single gravitational field is sometimes called `universal coupling'. Because of it, one can discuss the metric as a property of space-time itself rather than as a field over space-time. This is because its properties may be measured and studied using a variety of different experimental devices, composed of different non-gravitational fields and particles, and \ldots the results will be independent of the device. Thus, for instance, the proper time between two events is characteristic of spacetime and of the location of the events, not of the clocks used to measure it. (Will 2001)
\end{quotation}

Is this reasoning cogent? The purpose of this brief paper is to raise some doubts that complement those found in Anderson (1999). In section 2, it is questioned whether the universality of the behaviour of rods and clocks is indeed a basic feature of the theory, with special reference to the case of accelerating clocks.  In section 3, a detour is taken through special relativity, in which the changing nature of Einstein's views about the explanation 
of length contraction and time dilation is brought out by looking at a curious passage
in his 1949 \textit{Autobiographical Notes}. The notion of ``measurement" as it pertains to the behaviour of rods and clocks is then critically analysed. In the final section 4, attention turns back to general relativity. It is stressed that 
the universal behaviour of rods and clocks, to the extent that it exists, is not a consequence of the form of the Einstein field equations, any more than the very signature of the metric is, or any more than the local validity of special relativity is. The notion that the metric field can be viewed as ``a property of space-time itself rather than as a field over space-time'' is based on a feature of the theory --- related to what Will refers to as universal coupling --- that arguably sits some distance away from the central dynamical tenets of the theory. And the notion may even be misleading.

\section{Clocks and their complications}

In the last sentence of the above quotation, Will does not explicitly indicate that he is talking about infinitesimally close events. Indeed, it seems natural to assume that his remark extends to the notion of proper time between any two events that are connected by a finite curve that is everywhere time-like, and which represents the world-line of a clock. Of course in this case the proper time is defined relative to the curve. But in so far as the time read by the clock is related in the usual way to the length of the curve defined by the metric (i.e. the integral of $ds$ along the curve, for the line element $ds^2 = g_{\mu\nu}dx^\mu dx^\nu$), it is taken to be a universal phenomenon, i.e. independent on the constitution of the clock.

For most purposes, physical clocks are designated `good' or `ideal' when they tick in step with the temporal parameter appearing in the fundamental equations associated with our best theories of the non-gravitational interactions. But such considerations usually involve freely moving clocks, and even then no clock smaller than the whole universe can act strictly ideally because the action of the rest of the universe on it cannot be entirely screened off. Furthermore, if a very good approximation to an ideal clock can be found in regions of space-time with weak gravitational fields, when the same clock is placed in a strong gravitational field this behaviour will generally not persist, as Anderson (1999) stressed. How strong the tidal forces have to be to cause a significant degree of disruption to the workings of the clock will of course generally depend on the constitution of the clock.

And then there is the matter of accelerating clocks, corresponding to non-geodesic time-like worldlines.
For a correlation of the kind mentioned above, between the reading of an accelerating clock and the length of its world-line, to hold, it must be the case that the effect of motion on the clock at an instant can only be related to its instantaneous velocity, and not its acceleration. This condition is commonly (and somewhat misleadingly) referred to as the \textit{clock hypothesis} in both special and general relativity.$^1$ The question to be addressed in the rest of this section is whether the clock hypothesis, to the extent that is satisfied by real clocks, is indeed a universal phenomenon in the above sense. To this end, let's consider first one of the great early experimental tests of general relativity.
 
The Pound-Rebka experiment (Pound and Rebka 1960a), involving the emission and absorption of gamma rays by Fe$^{57}$ nuclei---using the newly-discovered M\"{o}ssbauer effect---placed at different heights in the Earth's gravitational field, is widely known for its role as corroborating Einstein's prediction of the general-relativistic red-shift effect.$^2$ It is perhaps less widely appreciated that the experiment also has a bearing on time dilation.  Both the emitter and absorber nuclei undergo accelerations due to thermal lattice vibrations, and Pound and Rebka (1960b), and independently Josephson (1960), had realized that even a small temperature difference between the emitter and absorber would result in an observable shift in the absorption line as detected in the Doppler shift method.$^3$ This shift, which must be taken into account in testing for the red-shift effect, is a consequence of the differential relativistic time dilation associated with the different root mean square (rms) velocities of the emitter and absorber nuclei. In fact, as was clarified in 1960 by Sherwin, the Fe$^{57}$ nuclei are playing 
the role of clocks in the ``twin paradox'', or ``clock retardation" effect. Note that the accelerations involved were of the order of 10$^{16}g$, and the accuracy of the experiment was within 10\%.$^4$ 

A later and more celebrated version of the retardation experiment was the 1977 CERN g-2 experiment (Bailey, Borer \textit{et al.} 1977; see also Wilkie 1977) with orbiting muons in a magnetic field and suffering accelerations of the order 10$^{18}g$.  A relativistic clock retardation effect was reported--- the clocks of course being the unstable muons --- to within an accuracy of 0.1\%.

In all such experiments, the clock retardation is calculated in conformity with the clock hypothesis, so that the effect is due to rms velocities or integration over the instantaneous velocities of the clocks: the instantaneous accelerations themselves are supposed to contribute nothing to the effect. But for any given clock, no matter how ideal its behaviour when moving inertially, there will in principle be an acceleration such that to achieve it the external force acting on the clock will disrupt its inner workings. As Eddington succinctly put it: ``We may force it into its track by continually hitting it, but that may not be good for its time-keeping qualities.'' (Eddington 1966). We can infer from the above experiments that Fe$^{57}$ nuclei and muons \textit{do} satisfy the clock hypothesis under accelerations of at least 10$^{16}g$ and 10$^{18}g$, respectively.  But this happy circumstance depends not on luck, nor definition, but on the physical make-up of the clocks in question.

In the case of the Pound-Rebka experiment, Sherwin provided a back-of-the-envelope calculation of the deformation of the Fe$^{57}$ nucleus caused by accelerations of 10$^{16}g$. Given the nature of the gamma ray resonance, the force associated with the relative displacement of the protons and neutrons within the nucleus is 3$\times 10^{28}$ dynes/cm. And given that only the protons are affected by the electric force, 
\begin{quotation}
\ldots the neutron should suffer a maximum relative displacement of about 1 part in $10^{13}$ of the nuclear diameter. Even using the great sensitivity of the M\"{o}ssbauer resonance, such a small distortion is not likely to produce an observable effect. First, it would have to produce a relative shift of the same order of magnitude between the two states which define the resonance. Second, the change in the resonance frequency arising from the acceleration would have to be independent of the direction of the acceleration, for, if it were not, the rapidly varying, cyclical acceleration patterns would have their effects averaged to zero over the emission time of a quantum (in a manner similar to that of the first-order Doppler shift arising from lattice vibrations). We conclude from this rough calculation that the mechanical distortion of the nuclear structure under the accelerations due to the lattice vibrations is very small, but under favorable circumstances an intrinsic acceleration-dependent effect in the resonance frequency might be observable. (Sherwin 1960)
\end{quotation}

In the case of the $g$-2 muon experiment at CERN, Eisele provided in 1987 a detailed analysis of the decay process for muons experiencing centripetal acceleration. With respect to inertial frames, such orbits are described in terms of a Landau-level with high quantum number, and Eisele used perturbation techniques in the theory of the weak interaction to calculate the approximate life time of the muons. He concluded that the correction to the calculation based on the clock hypothesis for accelerations of 10$^{18}g$ would be less than 1 part in 10$^{25}$, many orders of magnitude less than the accuracy of the 1977 experiment. Eisele also noted that near radio-pulsars, magnetic fields plausibly exist which could lead to an acceleration-induced correction to muon decay of almost 1\%. But he correctly concluded that 
\begin{quotation}
\ldots the most interesting part of this calculation surely consists not in any possible application like this but rather in the possibility in principle to verify the clock hypothesis in this special case [the $g$-2 CERN experiment] with the help of an accepted physical theory. (Eisele 1987)
\end{quotation}

Something important is going on here. Unlike time dilation induced by uniform motion, which normally is understood to be independent of the constitution of the clock, the effects of acceleration will depend on the magnitude of the acceleration and the constitution of the clock. 
Fe$^{57}$ nuclei and muons are much less sensitive to accelerations than `mechanical' clocks (like pendulum clocks), and the calculations of Sherwin and Eisele tell us why. They tell us why these microscopic clocks are capable of acting as hodometers, or `waywisers'$^5$ of time-like curves in relativistic space-time, even non-geodesic curves involving 3-accelerations of at least 10$^{16}g$. But they also remind us that the general validity of the clock hypothesis for such accelerations is not a forgone conclusion. (For an interesting example of a quantum clock failing to satisfy the clock hypothesis, see Knox 2008.) 

Before further discussion of the role of clocks in general relativity, it may be useful to remind ourselves of the origins of length contraction and time dilation in special relativity, without loosing sight of the fact that special relativistic effects are ultimately grounded in general relativity.

\section{Special relativity}

\subsection{Einstein's second thoughts}

In special relativity
it is clearly part of the theory that \textit{uniformly moving} rods and clocks contract and dilate respectively in a fashion that does not depend on their constitution. Although it is a very remarkable part of the theory, this phenomenon of universality tends to be taken for granted today. It was far less obvious when in the aftermath of the 1887 Michelson-Morley experiment, variants of the experiment were performed using different materials for the rigid structure supporting the optical equipment (with the result of course that no difference in the null outcome was observed). In Einstein's 1905 paper, this universality is simply assumed---it is built into the very operational significance he gave to the inertial coordinates and in particular their transformation properties. In addressing the question as to whether an explanation is actually available for the phenomonen, it is worth recalling the changing nature of Einstein's own explanation of the ``kinematical'' phenomena of length contraction and time dilation by way of the Lorentz transformations.

There is an oddity in Einstein's 1949 \textit{Autobiographical Notes} (Einstein 1969) concerning his reconstruction of the development of special relativity. Einstein stressed, as he did throughout his life, the role that thermodynamics played as a methodological template in his 1905 thinking leading to \textit{On the electrodynamics of moving bodies}. For reasons that have been much discussed in the literature, Einstein chose to base his new theory of space and time on principles which expressed ``generalizations drawn from a large amount of empirical data that they summarize and generalize without purporting to explain'' (Stachel 1998, p. 19) --- in other words, principles like the phenomenological laws of thermodynamics. One of these principles was of course the principle of relativity, which in Einstein's hands was restored to the \textit{universal} version originally defended by Galileo and Newton. (The nineteenth century had seen the rise of ether theories of light and later of electromagnetism more generally, which raised widespread doubts as to whether the classical relativity principle should strictly apply to the laws of electrodynamics. Einstein's aim was to banish such doubts, as Poincar\'{e} had actually urged before him, and even Lorentz and Larmor for up to second order effects.) 
Now it is important to emphasize that the formulation of this principle in 1905 does not presuppose the form of the (linear) coordinate transformations between inertial frames. After all, any hint that the Lorentz transformations are bound up in the expression of any one of Einstein's postulates would have opened up the derivation in the kinematical part of the 1905 paper to the charge of circularity. 

But note what Einstein says in 1949 recollections. In relation to the general problem of describing moving bodies in electrodynamics, he wrote:
\begin{quote}
By and by I despaired of the possibility of discovering the true laws by means of constructive efforts based on known facts. The longer and more despairingly I tried, the more I came to the conviction that only the discovery of a universal formal principle could lead us to assured results. The example I saw before me was thermodynamics. The general principle was there given in the theorem$^6$: 
the laws of nature are such that it is impossible to construct a perpetuum mobile (of the first and second kind). How, then, could such a universal principle be found? (Einstein 1969, p. 53.)
\end{quote}
The answer appears to be given, if at all, some paragraphs later:
\begin{quote}
The universal principle of the special theory of relativity is contained in the postulate: The laws of physics are invariant with respect to the Lorentz-transformations \ldots This is a restricting principle for natural laws, comparable to the the restricting principle of the non-existence of the \textit{perpetuum mobile} which underlies thermodynamics. (Einstein 1969, p. 57)
\end{quote}
What is striking here is that Einstein is seemingly conflating the main consequence of his 1905 postulates (the principle of Lorentz covariance of the fundamental laws of physics) with one of the postulates themselves, namely the phenomenological relativity principle, the principle so beautifully brought out in Galileo's famous thought experiment of the moving ship in his \textit{Dialogue concerning two chief world systems}, and expressed more succinctly in Corollary V of the Laws in the Newton's \textit{Principia}.$^7$ The principle of Lorentz covariance clearly has a much more awkward affinity with the laws of thermodynamics than the relativity principle. It (Lorentz covariance) is hardly a generalization drawn from a large amount of empirical data, especially in the context of 1905. 

Why this lapse on Einstein's part? I wonder if it was not because of the misgivings he had about the way he formulated his 1905 paper, misgivings which grew throughout his life. First, there is little doubt that right from the beginning he was aware of the limited explanatory power of what he called ``principle theories'' like thermodynamics. Secondly, when he confessed in 1949 to having committed in 1905 the ``sin'' of treating rods and clocks as primitive entities, and not as ``moving atomic configurations'' subject to dynamical analysis, he was merely repeating a point of self-correction he made in 1921. Finally, it is fairly clear that Einstein was increasingly unhappy with the central role that electrodynamics, and in particular the behaviour of light, played in his 1905 paper.

This last aspect of Einstein's reasoning brings us to the main point of this subsection. Einstein wrote in 1935:
\begin{quote}
The special theory of relativity grew out of the Maxwell electromagnetic equations. \ldots [but] the Lorentz transformation, the real basis of special-relativity theory, in itself has nothing to do with Maxwell theory. (Einstein 1935).
\end{quote}
Similarly, in a 1955 letter to Born, Einstein would write that the ``Lorentz transformation transcended its connection with Maxwell's equations and has to do with the nature of space and time in general''. He went on to stress that ``the Lorentz-invariance is a general condition for any physical theory.'' (Born \textit{et al}. 1971, p. 248). What is clear is that for the mature Einstein, the principle of Lorentz covariance, which applies to all the non-gravitational interactions, not just electrodynamics, is the heart of special relativity.$^8$ In stressing this point, Einstein was distancing himself from his formulation of 1905 with its emphasis on fundamental phenomenological postulates (one of which being the ``constancy'' of the speed of light relative to the ``rest'' frame). The principle of Lorentz covariance is still a restriction on fundamental laws, but it is not quite like any of the laws of thermodynamics. By putting emphasis on its primacy, Einstein is effectively saying that the phenomenological relativity principle is a consequence of something deeper.

\subsection{The sense in which rods and clocks don't ``measure''}

This last point was brought out nicely in the 1976 pedagogical work of John S. Bell (Bell 1976) in special relativity, which consisted essentially of an extension of Oliver Heaviside's 1889 result that the electrostatic field generated by a distribution of charge undergoes a distortion, according to Maxwell's equations, when the charge is put into motion. Heaviside's result was the inspiration for G. F. FitzGerald's speculation concerning the deformation of rigid bodies moving through the luminiferous ether, and Bell himself took inspiration from both Heaviside and the work of Joseph Larmor to calculate the effect of (slow) accelerative motion on a 2-dimensional atom consisting of a heavy positively charged particle being orbited by a negatively charged particle and modeled using Maxwell's equations, the Lorentz force law and the relativistic formula linking the moving particle's momentum with its velocity. Bell showed that the atom spatially contracts and its period dilates when it achieves a uniform speed, in accordance with relativistic predictions.$^9$ But more importantly for our purposes, he demonstrated that there is a new system of variables associated with the moving atom in relation to which the atom is described in exactly the same way that the stationary atom was described relative to the original variables --- and that the new coordinates were related to the original ones by a Lorentz transformation. Bell had effectively derived the relativity principle --- or that application of it to the simple electrodynamic system under consideration --- from the dynamics postulated to hold in the original frame. 

Bell of course realized that a satisfactory version of his atomic model would need to be reformulated in quantum theoretical terms, and that at root all the work in the argument is being done by the principle of Lorentz covariance of the fundamental equations. But what is important is that Bell's ``Lorentzian pedagogy'' is entirely free of the sin of treating rods and clocks as primitive entities,
 and it does not regard the relativity principle as fundamental. And in using this pedagogy as a warning against ``premature philosophizing about space and time'', Bell was reminding us that the reason rods and clocks do what they do is not because of what they are moving through but because of the dynamical principles of their very constitution. (Ironically, this lesson was in part lost to the ether theorists like Heaviside, FitzGerald and Larmor who were so influential on him.)

Bell also realized that the principle of Lorentz covariance needs to be assumed for \textit{all} the interactions governing the constitution of matter, not just the electromagnetic forces.
 Although he did not stress it, it is therefore a consequence of the approach he is advocating (as it was in the work of Swann in 1941, who had already applied the principle in the context of quantum theory) that rods and clocks contract and dilate respectively independently of their constitution. If Einstein had started with the universal principle of Lorentz covariance, he would not have needed to assume the universal nature of kinematics---that the inertial coordinate transformations codify the behaviour of rods and clocks whatever they are made of. 

This point leads us to consider the question as to what it means to say such entities ``measure''. Much work has been expended in quantum theory in order to specify what is meant by a measurement process, inspired in good part by the so-called `measurement problem' --- the problem of accounting, solely within quantum theory itself, for the emergence of well-defined results in standard measurement procedures. Models of such procedures are given by specifying a certain interaction Hamiltonian governing the coupling between the microscopic `object' system and the (usually macroscopic) measurement device. As a result of this coupling, the final state of the apparatus becomes correlated, in the appropriate sense, with the initial state of the object system, the combined system being subject to the time-dependent Schr\"{o}dinger equation. Nothing like \textit{this} notion---correlation through coupling---is taking place when rods and clocks do what they do in relativity theory, either at rest or in motion.

Bell's 1976 message was a (probably unwitting) reiteration of the message from a number of earlier commentators. Even before Swann, Pauli, for example, had already stressed in 1921, in his magisterial survey of relativity theory (Pauli 1921), that moving rods and clocks would not contract and dilate in conformity with the special relativistic predictions, were it not for the fact that all the non-gravitational interactions which account for the cohesive forces between the micro-constituents of these bodies are governed by Lorentz covariant equations. These objects are not measuring anything in the above strong sense; they are merely acting in accordance with the dynamical laws governing their internal make-up. They are not the analogue of thermometers of the space-time metric. They measure  in the weaker sense that their behaviour \textit{correlates} with aspects of space-time structure and that is why we are interested in them and use our best theories to build them$^{10}$ --- but it is not because space-time \textit{acts} on them in the way a heat bath acts on a thermometer, or the way a quantum system acts on a measuring device.

\section{General relativity}

This last claim might be regarded as much more contentious in general relativity (GR) than in special relativity. After all, in special relativity the Minkowski metric is absolute and non-dynamical.$^{11}$ But in GR, the action-reaction principle seems gloriously restored, its reincarnation now involving matter and metric. But how does general relativity purport to predict that an ideal clock, for example, will act as a way-wiser of space-time without treating it as a primitive entity?$^{12}$ 

It may be useful in answering this question to consider momentarily the case of alternative theories of gravity which feature more than one metric field, and in particular \textit{bimetric} theories. For example, Rosen (1980), in an attempt to avoid the singularities that appear in standard GR, introduced besides the $g_{\mu\nu}$ field that describes gravity, a non-dynamical metric field $\gamma_{\mu\nu}$ of constant curvature which serves to define a fundamental rest frame of the universe. More recently, Bekenstein (2004) has developed a bimetric theory which is a covariant version of Milgrom's MOND program, designed principally to account for the anomalous rotation curves in galaxies without appeal to dark matter. Bekenstein's Tensor-Vector-Scalar theory (T\textit{e}V\textit{e}S) incorporates two metric fields.$^{13}$ The first, represented by $g_{\mu\nu}$, has as its free Lagrangian density the usual Hilbert-Einstein Lagrangian. The second, represented by $\tilde{g}_{\mu\nu}$, can be expressed as a deformation --- a disformal transformation --- 
of $g_{\mu\nu}$ according to a formula that depends on fundamental vector and scalar fields postulated to exist alongside $g_{\mu\nu}$. What is of interest to us in the present context is not whether these bimetric theories are true, 
but how it is in each theory that the usual rods and clocks end up surveying (at most) only one of the two metrics, as they must.

Take T\textit{e}V\textit{e}S. It is critical to the enterprise that the metric which is assigned chronometric significance is the ``less basic'' $\tilde{g}_{\mu\nu}$, not $g_{\mu\nu}$. It is $\tilde{g}_{\mu\nu}$ that is ``measured'' by rods and clocks, and whose conformal structure is traced out by light rays and whose time-like geodesics are the possible worldlines of free bodies. Bekenstein is clear as to how this is realized: $\tilde{g}_{\mu\nu}$ is ``delineated by matter dynamics'' in the right way. The issue has to do with the way the usual matter fields are postulated to couple to $\tilde{g}_{\mu\nu}$ (and therefore to $g_{\mu\nu}$). In short, the \textit{Einstein equivalence principle} is defined relative to $\tilde{g}_{\mu\nu}$ in the theory. Bekenstein, in establishing the operational significance of $\tilde{g}_{\mu\nu}$ in T\textit{e}V\textit{e}S, is simply doing what is done --- though often with less emphasis --- in relation to $g_{\mu\nu}$ in standard GR.

Let us consider an arbitrary event $P$ in space-time. There exist in the neighbourhood of $P$ locally inertial (Lorentz) coordinates, such that at $P$ the first derivatives of $g_{\mu\nu}$ vanish and $g_{\mu\nu} = \eta_{\mu\nu}$ (read $\tilde{g}_{\mu\nu}$ for $g_{\mu\nu}$ in T\textit{e}V\textit{e}S). The Einstein equivalence principle has two components. The first (universality) states that the fundamental laws for \textit{all} the non-gravitational interactions involving matter fields take their simplest form at $P$ relative to these local coordinates. The second (minimal coupling) states that this form is the special relativistic form; in particular neither the Riemann curvature tensor nor its contractions appear in the laws. As Ohanian put the principle: ``At each point of space-time it is possible to find a coordinate transformation such that the gravitational field variables can be eliminated from the field equations of matter.'' (Ohanian 1977). It ensures, in the words of Misner, Thorne and Wheeler, that ``... in any and every local Lorentz frame, anywhere and anytime in the universe, all the (nongravitational) laws of physics must take on their familiar special relativistic forms.''  (Misner, Thorne and Wheeler 1973, p. 386.)

It follows from the Einstein equivalence principle that the fundamental equations governing all the non-gravitational interactions are locally Lorentz covariant$^{14}$, and hence that in so far as we can ignore the effects of tidal forces on rods and clocks, they will behave in conformity with the predictions of special relativity, as stressed in section 3. (In the case of accelerating rods and clocks, the kinds of qualifications raised in section 2 will of course still be relevant to the issue of the universality of their behaviour.) And so the phenomenological relativity principle (or rather its local variant) for the non-gravitational interactions is itself a consequence of the Einstein equivalence principle.

In stressing the role of this principle in accounting for the behaviour of rods and clocks in general relativity, an observation due to James L. Anderson in his remarkable 1976 book \textit{Principles of Relativity Physics} comes to mind. Anderson claims that of the two components above of the principle, the first (which implies that measurements on any physical system will determine the same affine connection) is essential to GR --- to any metric theory of gravity --- and the second is not. (Anderson 1967, section  10-2.) Be that as it may, it is worth recalling that the Einstein equivalence principle is not a strict consequence of the form of Einstein's field equations, any more than the signature of the metric is, or any more than the Galilean covariance of Newtonian mechanics follows from the strict form of Newton's laws of motion. (The further assumptions that all Newtonian forces are velocity-independent like the gravitational force, and similarly that inertial masses are velocity-independent, are jointly  required for the Galilean covariance of Newton's laws.) If the metric field is to be considered a property of space-time, in the sense of Will above, it requires a very non-trivial dynamical assumption to be made over and above postulating the field equations, which is roughly that special relativity, properly understood, holds locally.

But just because the metric field \textit{can} be so considered, it is not clear it \textit{should} be. Or rather, it is not clear that the geometrical interpretation of $g_{\mu\nu}$ is intrinsic to its dynamical role in GR, particularly when one considers a non-trivial space-time completely free of matter and hence rods and clocks. Even in the general case, the notion that the $g_{\mu\nu}$ field is the fabric of space-time, rather than a field in space-time, may be misleading.$^{15}$ It may serve to hinder recognition of the possibility that Einstein gravity is an emergent phenomenon, for example in the sense that it is a consequence of the specific dynamics of an evolving fundamental 3-geometry$^{16}$, or in the stronger sense  that the field equations are analogous to the equations of fluid dynamics, which emerges from molecular physics as a low-momentum long-distance approximation.$^{17}$ These non-standard approaches may not prove to be correct, but they---and particularly the latter view which goes some way to deriving the Lorentzian signature of the metric, and furthermore calls into question the program of quantizing gravity---should not be dismissed lightly. At the very least, the view that $g_{\mu\nu}$ \textit{is} space-time structure may serve to hide from view the fact that the Einstein equivalence principle is a highly non-trivial part of GR, and that the universality of the principle of Lorentz covariance incorporated threrein is arguably mysterious---by which I mean explanation-seeking---particularly in the absence of a strict theoretical unification of all the non-gravitational interactions.$^{18}$ 

It is noteworthy that for whatever reasons, Einstein at the end of his life sounded clear notes of caution on the question of the interpretation of the metric field. In 1948, he wrote in a letter to Lincoln Barnett:
\begin{quote}
I do not agree with the idea that the general theory of relativity is geometrizing Physics or the gravitational  field. The concepts of Physics have always been geometrical concepts and I cannot
see why the $g_{ik}$ field should be called more geometrical than f.[or]
i.[nstance] the electro-magnetic field or the distance of bodies in
Newtonian Mechanics. The notion comes probably from the fact
that the mathematical origin of the $g_{ik}$ field is the Gauss-Riemann
theory of the metrical continuum which we are wont to look at
as a part of geometry. I am convinced, however, that the distinction
between geometrical and other kinds of fields is not logically
founded.$^{19}$
\end{quote}
And in his 1949 \textit{Autobiographical Notes}, when attempting to justify his ``sin'' of treating rods and clocks as primitive in his 1905 relativity paper by appealing to the (then) lack of understanding of the microphysics of matter, he stressed:
\begin{quotation}
But one must not legalize 
the mentioned sin so far as to imagine that intervals are physical entities of a special type, intrinsically different from other physical variables (``reducing physics to geometry'', etc.). Einstein (1969), p. 61.
\end{quotation}

\section{Acknowledgments}

I wish to thank David Rowe for the kind invitation to contribute to this volume. I am grateful to him and Dennis Lehmkuhl for helpful comments on the first draft of this paper. I thank Norbert Straumann for bringing to my attention, during the \textit{Beyond Einstein} conference in Mainz, the 1987 work by Anton Eisele, and for further correspondence. I  also benefitted from discussions with Eleanor Knox, David J. Miller, Constantinos Skordis and George Svetlichny, as well as Adrian Sutton and his project students (see note 9).
		
\section{Endnotes}

\noindent 1. For a discussion of the role of the clock hypothesis within special and general relativity, see (Brown 2005, section 6.2.1).

\noindent 2. It is noteworthy that this famous Harvard experiment was not the first to test the red-shift hypothesis; it was preceded in 1960 by a similar experiment using the M\"{o}ssbauer effect performed at Harwell in the UK (Cranshaw, Schiffer \textit{et al.} 1960). The Harwell group also performed in the same year another red-shift test, again using the M\"{o}ssbauer effect, but this time involving a source at the centre of a rotating wheel which contained a thin iron absorber (Hay, Schiffer \textit{et al.} 1960; see also Sherwin 1960). 

In 1960 it was clarified by Schild that that red-shift tests are in their nature insensitive to the precise form of Einstein's field equations (Schild 1960). In the same paper, he also noted that nonetheless the red-shift phenomenon itself leads naturally to the idea that gravitational fields are related to the curvature of space-time. It is worth emphasizing that this argument for curvature depends on appeal to the inhomogeneity of the gravitational field and therefore to a series of red-shift experiments sufficiently separated on the surface of the Earth; a single example of the red-shift, such as in the Pound-Rebka experiment, in which the equipment is largely insensitive to tidal forces, is not enough. This point is sometimes overlooked, as in Carroll's otherwise excellent (2004), pp. 53-4.

\noindent 3. If, for example, the source is at a higher temperature than the absorber, the shift is negative. So for resonance absorption to occur in this case, the absorber must be given a small velocity away from the source so that the Doppler effect can compensate for the shift. See Sherwin (1960), p. 19. 

\noindent 4. An earlier 1960 redshift test, also involving the M\"{o}ssbauer effect, involved a source at the centre of a rotating wheel which contained a thin iron absorber (Hay, Schiffer \textit{et al.} 1960; see also Sherwin 1960). Again a frequency shift due to relativistic time dilation---a case of clock retardation---was detected to an accuracy of a few percent; the radial acceleration in this experiment was of the order of 10$^4g$.
		
\noindent 5. For the relevance of the term `waywiser', see Brown (2005), pp. 8 and 95. The above calculation given by Sherwin in relation to the iron nuclei in the Pound-Rebka experiment answers, indeed anticipates, the point recently made by Lyle (2010, section 1.1.10), that a dynamical proof needs to be provided of the claim that the iron nuclei in this experiment satisfy the clock hypothesis to a good approximation.

\noindent 6. It is noted in Brown (2005, p. 71, footnote 8) that the word `theorem' might be more happily translated from the original German as `statement'.

\noindent 7. In his 1949 discussion, Einstein clearly appreciates the difference between the two principles; see (Einstein 1969), p. 57.	

\noindent 8. In 1940, Einstein wrote: ``The content of the restricted relativity theory can accordingly be summarised
in one sentence: all natural laws must be so conditioned that they
are covariant with respect to Lorentz transformations.'' (Einstein 1954, p. 329) It is worth recalling in this context the way Einstein described in his \textit{Autobiographical Notes} the main contribution Minkowski made to relativity theory. It was not so much Minkowski's ontological fusion of space and time into a single four-dimensional entity that Einstein praised, but his provision of a tensor calculus in which equations for the non-gravitational interactions are manifestly Lorentz covariant. For Einstein, Minkowski had done for relativity what Heaviside and others did for Maxwell theory when they introduced the three-vector formulation of electrodynamics (so that the physics is manifestly Euclidean covariant). Minkowski ``showed that the Lorentz transformation \ldots is nothing but a rotation of the coordinate system in the four-dimensional space'' (Einstein 1969, p. 59), an insight which in fact Poincar\'{e} had anticipated.

\noindent 9. Suppose one considers the possibility of modeling a rigid rod by way of an infinite crystal composed of ions held together by electrostatic forces, rather in the spirit of Lorentz's 1892 model of a system of charges held together in unstable equilibrium. Then the dynamical analysis seems to lead (as it did in the Lorentz case; see Brown 2001) to a certain motion-induced deformation, rather than a strict longitudinal contraction: the conformal covariance of the equations has not been broken. This point was brought home to me in recent discussions with Adrian Sutton and his 4th year undergraduate project students in the Department of Physics at Imperial College London: H. Anwar, V. Venkataraman, A. Wiener, C. Chan, B. Lok, C. Lin, and G. Abdul-Jabbar. This group has been studying a constructive approach to length contraction, similar to that of Bell (1976), but in which the effects of motion are calculated in a classical model of the attractive interatomic forces in an infinite ionic crystal. The question that has been thrown up, as I see it, is whether in such models it is possible to obtain strictly longitudinal length contraction without introducing quantum mechanics. (In the Bell atomic model, the conformal symmetry is broken by appeal to the Lorentz force law for the orbiting charge.) This question also applies to another electrostatic model of a rigid rod provided by Miller (2009), which was brought to my attention after discussions with the Imperial College group. Here, the way the author effectively breaks the conformal symmetry in the electrodynamics is not entirely consistent with the `constructive' nature of his approach.

\noindent 10. Note that the distinction here between strong and weak measurement is not relevant to the issue of the accuracy of the measurement.

\noindent 11. A possible objection to this reasoning might go as follows. The generally covariant formulation of any specially relativistic dynamics (such as Maxwellian electrodynamics), or more generally any dynamical theory within an absolute space-time background, flat or curved, leads to equations of motion in which the absolute structure appears in the equations. Such structure appears to be causally relevant; indeed a violation of the action-reaction principle seems to obtain. Space-time structure acts on matter, but not the other way round. However,
	demanding general covariance in the context of special relativity is like demanding that (`pure gauge') electromagnetic vector and scalar potentials appear in the Schr\"{o}dinger equation for a free particle. Just as it would be odd to say that such potentials are physically acting on the particle, arguably the `action' of space-time structure in special relativity is merely an artifact of the generally covariant formulation, which is ill-suited to the theory. (For further discussion of the purported violation of the action-reaction principle in special relativity theory, see Brown and Pooley (2004) and Brown (2005), section 8.3.1.) 
	
\noindent 12. It is curious how infrequently this issue is raised. A rare case was Dieks, writing in 1987: ``... it should be emphasized that the general theory of relativity is a fundamental physical theory ... it can safely be said that constructs like macroscopic measuring rods and clocks cannot figure as essential elements in such a fundamental theory. ... [T]he behaviour of macroscopic bodies like rods and clocks should be explained on the basis of their microscopic constitution.'' (Dieks 1987, p. 15; see also Dieks 1984.) However, the nature of this explanation as suggested by Dieks differs from what follows.

\noindent 13. For a recent review of T\textit{e}V\textit{e}S, see Skordis (2009). It has been argued (Zlosnik \textit{et al}. 2006) that T\textit{e}V\textit{e}S is not a true bimetric theory. First, it can be shown to be equivalent to a (mathematically more complicated) Tensor-Vector theory involving just the single metric  $\tilde{g}_{\mu\nu}$ in the total action. More significantly, these authors claim that tensor gravity waves propagate along the same light cone as electromagnetic ones. But this claim conflicts with the analysis of T\textit{e}V\textit{e}S and its generalizations by Skordis (2006, 2008, and 2009).

\noindent 14. A gravitational theory that violates local Lorentz covariance is due to Jacobson and Mattingly (2001). It contains a time-like unit vector field which serves to pick out a preferred frame. I take it the appearance of this field in the equations governing the matter fields is ruled out by the second component of the Einstein equivalence principle.

\noindent 15. A different analysis leading to the same conclusion was provided by James L. Anderson in his (1999).  Anderson also argued that the (approximately) metrically-related behaviour of clocks can be derived from the dynamical assumptions of GR, in the same way that the motion of a free test particle can be derived. He claimed that this becomes particularly clear in the approximation scheme developed to address the problem of motion in GR due to Einstein, Infeld and Hoffmann in 1939 and 1940. I repeat Anderson's concluding remarks:
\begin{quote}
In this paper I have argued that a metric interpretation is not needed in general relativity and that the purposes
for which it was originally introduced, i.e., temporal and spatial measurements and the determination of geodesic
paths, can be all be derived from the field equations of this theory by means of the EIH [Einstein, Infeld and Hoffman] approximation scheme. As a
consequence, the only \textit{ab initio} space-time concept that is required is that of the blank space-time manifold. In this
view what general relativity really succeeded in doing was to eliminate geometry from physics. The gravitational field
is, again in this view, just another field on the space-time manifold. It is however a very special field since it is needed
in order to formulate the field equations for, what other fields are present and hence couples universally with all other
fields.
\end{quote}

\noindent 16. See the reconstruction of GR due to Barbour, Foster and \'{O} Murchadha (Barbour \textit{et al.} 2002) based on a dynamical 3-geometry approach and inspired by Mach's relational reasoning.

\noindent 17. The emergent approach discussed by Barcel\'{o} \textit{et al.} (2001) relies on classical considerations related to the so-called ``analog models'' of GR to motivate the existence of the Lorentzian metric field, and effective theories arising out of the one-loop approximation to quantum field theory to generate the dynamics (and in particular the familiar Hilbert-Einstein term in the effective action) in the spirit of Sakharov's 1968 notion of induced gravity. Such an approach clearly calls into question the appropriateness of quantizing gravity. For more recent developments along similar lines, but now based on a potentially deep connection between the field equations for the metric and the thermodynamics of horizons, see Padmanabhan (2007, 2008).

\noindent 18. The above-mentioned 2002 work of Barbour \textit{et al.} was originally thought to provide a derivation of the Einstein equivalence principle. However, careful further analysis by Edward Anderson has cast doubt on this claim; for details see Anderson (2007). It should be mentioned that the validity of the Einstein equivalence principle seems to be genuinely mysterious in the case of the case of the emergent gravity approach; see Barcel\'{o} \textit{et al.} (2001), section 4. 

\noindent 19. I am grateful to Dennis Lehmkuhl for recently bringing this letter to my attention. It is reproduced in the preface by John Stachel in Earman, Glymour \textit{et al.} (1977), p. ix. A fuller account of most of the main arguments made in sections 3 and 4 of the present paper is found in Brown (2005).

\section{References}

Anderson, Edward (2007). ``On the recovery of geometrodynamics from two
different sets of first principles'', \textit{Studies in History and Philosophy of
Modern Physics}  \textbf{38}, 15-57.

Anderson, James L. (1967). \textit{Principles of Relativity Physics},  New York: Academic Press Inc.  

Anderson, James L. (1999). ``Does general relativity require a metric'', arXiv:gr-qc/9912051v1.     

Bailey, J., K. Borer, F. Combley, H. Drumm, F.J.M. Farley, J.H. Field, P.M. Hattersley, F. Krienen, F. Lange, E. Picasso and W. von R\"{u}den (1977).
``Measurements of Relativistic Time Dilation for Positive and Negative Muons in a Circular Orbit", \textit{Nature} \textbf{268}, 301-305. 

Barcel\'{o}, Carlos, Matt Visser and Stefano Liberati (2001). ``Einstein gravity as an emergent phenomenon?'', arXiv:gr-qc/0106002v1.

Bekenstein, Jacob D. (2004). ``Relativistic gravitation theory for the modified Newtonian dynamics paradigm'', \textit{Physical Review D}, \textbf{70}, 083509.

Bell, John. S. (1976). ``How to teach special relativity'', \textit{Progress in Scientific Culture}, \textbf{1}, reprinted in J.S. Bell, \textit{Speakable and Unspeakable in Quantum Mechanics}, Cambridge: Cambridge University Press (2008), pp. 67-80.

Born, Max, H. Born and A. Einstein (1971). \textit{The Born-Einstein Letters}, London: Macmillan.

Brown, Harvey R. (2001). ``The origins of length contraction: I the FitzGerald-Lorentz deformation hypothesis'', \textit{American Journal of Physics} \textbf{69}, 1044-1054 (2001); arXive:gr-qc/0104032; PITT-PHIL-SCI 218.

Brown, Harvey R. (2005). \textit{Physical Relativity. Space-time structure from a dynamical perspective}, Oxford: Oxford University Press. The 2007 paperback edition removes a number of errata from the original.

Brown, Harvey R. and Pooley, Oliver (2004). ``Minkowski space-time: a glorious non-entity'', arXiv:gr-qc/010004032; PTT-PHIL-SCI 1661. A revised version appeared in \textit{The Ontology of Space, 1}, D. Dieks (ed.), Amsterdam: Elsevier, 2006, pp. 67-89.

Carroll, Sean (2004). \textit{Spacetime and Geometry. An introduction to General Relativity},  San Francisco: Addison-Wesley.

Cranshaw, T.E., J.P. Schiffer and A.B. Whitehead (1960). ``Measurement of the Gravitational Red Shift Using the M\"{o}ssbauer Effect in Fe$^{57}$'', \textit{Physical Review Letters} \textbf{4}, 163-4.

Dieks, Dennis (1984).  ``On the Reality of the Lorentz Contraction'', \textit{Zeitschrift f\"{u}r allgemeine Wissenschaftstheorie}, \textit{15}, 330-42.

Dieks, Dennis (1987). ``Gravitation as a universal force'', \textit{Synthese}, \textbf{73}, 381-39.

Earman, John, Clark Glymour and John Stachel (eds.) (1977). \textit{Minnesota Studies in the Philosophy of Science, Volume VIII: Foundations of Space-Time Theories (proceedings)}, Minnesota: University of Minnesota Press.

Eddington, Arthur S. (1966). \textit{Space, Time and Gravitation. An Outline of the General Theory of Relativity}, Cambridge: Cambridge University Press.

Einstein, Albert (1935). ``Elementary derivation of the equivalence of mass and energy'', \textit{Bulletin of the American Mathematical Society}, \textbf{41}, 223-30.

Einstein, Albert (1954). ``The Fundamentals of Theoretical Physics'', in his \textit{Ideas and Opinions}, New York: Bonanza Books; pp. 323-335. 

Einstein, Albert (1969). ``Autobiographical Notes'', in P.A. Schilpp (ed.), \textit{Albert Einstein: Philosopher-Scientist, Vol 1}, Illinois: Open Court; pp. 1-94.

Eisele, Anton (1987). ``On the behaviour of an accelerated clock'', \textit{Helvetica Physica Acta} \textbf{60}, 1024-1037.

Hay J.J., J P Schiffer, T E Cranshaw and P A Egelstaff (1960).  ``Measurement of the red shift in an accelerated system using the M\"{o}ssbauer effect in Fe$^{57}$'', \textit{Physical  Review Letters} \textbf{4}, 165-166.

Jacobson, Ted and and James Mattingly (2001). ``Gravity with a dynamical preferred frame'', \textit{Physical Review D}, \textbf{64}, 024028.

Josephson, B.D. (1960). ``Temperature-dependent shift of $\gamma$ rays emitted by a solid'', \textit{Physical Review Letters}, \textbf{4}, 341-2.

Knox, Eleanor (2008). ``Flavour-Oscillation Clocks and the Geometricity of General Relativity'', arXiv:0809.0274v1.

Lyle, Stephen (2010). ``Rigidity and the Ruler Hypothesis'', to appear in V. Petkov (ed.), \textit{Space, Time, and Spacetime - Physical and Philosophical Implications of Minkowski's Unification of Space and Time}, Berlin, Heidelberg, New York: Springer, 2010; and almost unchanged in S. Lyle, \textit{Self-Force and Inertia. Old Light on New Ideas.} Lecture Notes in Physics; Berlin, Heidelberg: Springer, 2010.

Miller, David J. (2009). ``A constructive approach to the special theory of relativity'', arXiv:0907.0902v1 [physics.class-ph].

Misner, Charles W., Kip S. Thorne and John A. Wheeler (1973). \textit{Gravitation}, San Francisco: Freeman \& Co.

Ohanian, Hans C. (1977). ``What is the principle of equivalence?'', \textit{American Journal of Physics}, \textbf{45}, 903-909.

Padmanabhan, T. (2007). ``Gravity as an emergent phenomenon: A conceptual description''. arXiv:0706.1654v1 [gr-qc].

Padmanabhan, T. (2008). ``Gravity: the inside story'', \textit{General Relativity and Gravitation} \textbf{40}, 2031-2036.

Pauli, Wolfgang (1921). ``Relativit\"{a}tstheorie'', \textit{Encyklop\"{a}die der matematischen Wissenschaften, mit Einschluss ihrer Anwendungen vol 5 Physik}, A. Sommerfeld (ed.); Leibzig: Tauber. English translation: \textit{Theory of Relativity}, New York: Dover, 1981.

Pound, R.V. and G.A. Rebka Jr. (1960a). "Apparent weight of photons", \textit{Physical Review Letters}, \textbf{4}, 337-41.

Pound, R.V. and G.A. Rebka Jr. (1960b). ``Variation with temperature of the energy of recoil-free gamma rays from solids'', \textit{Physical Review Letters} \textbf{4}, 274-5.

Rosen, Nathan (1980). ``General relativity with a background metric'', \textit{Foundations of Physics}, \textbf{10}, 637-704.

Schild, Alfred (1960). ``The Equivalence Principle and Red-Shift Measurements'', \textit{American Journal of Physics} \textbf{28}, 778-780.

Sherwin, C.W. (1960). ``Some Recent Experimental Tests of the `Clock Paradox' '', \textit{Physical Review} \textbf{120}, 17-24.

Skordis, Constantinos (2006). ``Tensor-vector-scalar cosmology: Covariant formalism for the background evolution
and linear perturbation theory'', \textit{Physical Review D}, \textbf{74}, 103513.

Skordis, Constantinos (2008). ``Generalizing tensor-vector-scalar cosmology'', \textit{Physical Review D}, \textbf{77},123502.

Skordis, Constantinos (2009) ``The Tensor-Vector-Scalar theory and its cosmology'', arXiv: 0903.3602v1

Stachel, John (1998), ``Introduction'', in \textit{Einstein's Miraculous Year. Five papers that changed the face of physics}, J. Stachel (ed.), Princeton: Princeton University Press; pp. 3-27.

Swann, W.F.G. (1941), ``Relativity, the Fitzgerald-Lorentz contraction, and quantum theory'',  \textit{Reviews of Modern Physics} \textbf{13}, 190-6.

Wilkie, Tom (1977). ``The Twin Paradox Revisited''. \textit{Nature} \textbf{268}, 295-296.

Will, Clifford M. (2001). ``The Confrontation between General Relativity and Experiment'', 
http://www.livingreviews.org/lrr-2001-4 

Zlosnik, T.G, P.G. Ferreira, and G.D. Starkmann (2006). ``The Vector-Tensor nature of Bekenstein's relativistic theory of Modified Gravity'', \textit{Physical Review D} 74:044037; arXiv:gr-qc/0606039v1.

\end{document}